# A CONTRIBUTION TO SECURE THE ROUTING PROTOCOL "GREEDY PERIMETER STATELESS ROUTING" USING A SYMMETRIC SIGNATURE-BASED AES AND MD5 HASH

Mohammed ERRITALI and Oussama Mohamed Reda and Bouabid El Ouahidi [1]

[1] Data mining and networks laboratory, Department of Computer Science, Mohamed V University – Faculty of Sciences Rabat – Morocco
`mederritali@yahoo.fr` , `oussama.reda@fsr.ac.ma` , `ouahidi@fsr.ac.ma`

## ABSTRACT

*This work presents a contribution to secure the routing protocol GPSR (Greedy Perimeter Stateless Routing) for vehicular ad hoc networks, we examine the possible attacks against GPSR and security solutions proposed by different research teams working on ad hoc network security. Then, we propose a solution to secure GPSR packet by adding a digital signature based on symmetric cryptography generated using the AES algorithm and the MD5 hash function more suited to a mobile environment.*

## 1. INTRODUCTION

The wireless communications networks are now ubiquitous in our daily life and offer many opportunities in many areas.
Vehicular ad hoc networks (VANETs) are special networks that s' appear in this context. These networks are formed according to the appearance and movement of vehicles and its consist of a set of moving objects witch communicate with each other using wireless networks like IEEE 802.11 and Ultra Wide Band (UWB) [1].
These communication systems between vehicles can be used to develop applications to improve road safety or allow Internet access for passengers. These applications are known as intelligent transportation systems (ITS) [1].
ITS applications can be both implemented in the infrastructure of roads and the vehicles themselves. This type of operation gives three types of ITS applications deployments [1]: vehicle-to-vehicle (V2V), infrastructure-to-vehicle (I2V) and vehicle-to-infrastructure (V2I). These ITS applications require robust communications a minimum quality of service and security when exchanging data. However, this contrasts greatly with the highly dynamic nature of vehicular networks (topology change, variable distance between vehicles, frequent loss of connectivity, unreliable communications, delay ...).
Security remains the weak link in these new networks, networks Vanet are by nature very vulnerable to various types of attack as unauthorized access, spoofing, eavesdropping, message editing and Denial of Service (DoS ). [2,3,15,16]
This fundamental problem of security does not lie only at the physical media but also in the fact that all nodes are equivalent and potentially necessary for the proper functioning of the network. In addition, lack of infrastructure for the authentication of nodes, a malicious or compromised node might fit and perform various malicious actions such as modification or message forgery.

The cons-measures to prevent such attacks often rely on cryptography, which enables encryption, generating a digest and / or digitally signing messages exchanged through the network.

These techniques can be implemented either by means of symmetric cryptography, with the same shared secret key to encrypt and to decrypt and uses hash functions to generate digests, or by means of asymmetric cryptography, two keys different pair (private key / public key) for encryption and decryption, and which assigns a different key signature to each participant. This latter mechanism requires the establishment of a Public Key Infrastructure (PKI) [8], in most cases the presence of a Certification Authority (CA) to certify that a certain key belongs to such user.

In this first part of our work we present an overview on possible attacks against GPSR [4] and security solutions proposed by different research teams working on ad hoc network security. The second part relates to our contribution to secure GPSR by adding a digital signature generated using a symmetric algorithm AES, which is to protect against attacks that target the routing protocol. Finally we present the cryptographic algorithms chosen and the experimental results.

## 2. VEHICULAR NETWORK SECURITY

Research on the safety of VANET is just beginning, with few papers and has not yet been developing or standardized a complete solution.

The Vanets networks are by nature more sensitive to the problems of safety. The intrusion on the support of transmission is easier conducting denial of service attack by jamming the frequency bands used.

The context of vehicle to vehicle communication also increases the number of potential security vulnerabilities. Being by definition without infrastructure, the vehicular ad hoc networks cannot profit from the services of safety offered by dedicated equipment: firewalls, authentication servers, these security services should be distributed and cooperative.

The routing also poses specific problems: each network station can serve as a relay and therefore the ability to capture or divert transit traffic. Denial of service attacks are also possible.

### 2.1. Attacks against routing in VANET

Attacks against the routing protocol for vehicular ad hoc networks can be designed to change the protocol itself, so that traffic passes through a node controlled by the adversary. An attack can also be designed to prevent the formation of the network, forcing the nodes to store incorrect routes, and generally disrupt the network topology.

Attacks at the routing can be classified into two categories: generation and incorrect relaying traffic [2].

### 2.1.1. Incorrect traffic generation

This includes attacks that consist of false signalling messages sent with the identity of another node (identity spoofing). The consequences are a possible conflict of information in different parts of the network, communications degradation and unreachable nodes [14].

In a geographical routing protocol, an adversary node may declare a position closer to all destinations so that all nodes around it will route their packets to node opponent. Then, the opponent can cut communications in the network by rejecting packets received instead of forward.

An adversary can either perform a Denial of Service by saturating the support with a large amount of broadcast messages, reducing the rate of knots and, at worst, preventing them from communicating

### 2.1.2. Incorrect relaying traffic

Communications from legitimate nodes can be contaminated by malicious nodes. A node opponent can avoid relaying the messages it receives in the end reduce the amount of information available to other nodes. This was called black hole attack (attack black hole) [5]. And this is a simple way to perform a DOS. This attack can be performed on all or a portion of the received packets, making it unavailable or difficult to reach the destination node.
An adversary can also modify the messages it receives before returning, if a system to ensure the integrity digest has not been established.

### 2.2. Secure routing protocols:

The most known protocols are: SRP [17], SAODV [12, 13] and ARIADNE [18].SRP requires a pre-existing security association between the source node and destination node. The authors propose a mechanism to detect malicious behaviour (Neighbour Lookup Protocol) and a mechanism for securing data (Secure Message Transmission Protocol), they propose to use them in complementary with SRP.
SAODV secures the modifiable data routing. The mechanism secures the main editable field hop count, which may for example be decremented by an attacker deliberately, and authenticates fields that should not be changed using a digital signature.
ARIADNE is used to authenticate routing messages using three mechanisms. The first use shared secrets between each pair of nodes, the second a combination of shared secret between the nodes communicate and disseminate authenticated (TESLA), the third uses digital signature. We inspired from the experience of these protocols to provide a secure solution for ad hoc routing protocols GPSR.

### 2.3. Protection of the routing protocol GPSR

When we talk about securing the routing, we want to ensure the integrity, non-repudiation and availability of service. The protection of routing messages is guaranted by a signature or digests. Indeed the lack of centralized infrastructure in vehicular ad hoc networks undermines the direct use of authentication systems based on public key cryptography. These authentication systems involve the use of certificates issued by a central authority.
There are three major trends in the field of authentication for wireless ad hoc networks. Two of these guidelines are based on the establishment of a secret key for subsequent authentication of participants.
The two models based on a secret key are:
   -The key agreement: Diffie-Hellman.
The Diffie-Hellman, named after its authors Whitfield Diffie and Martin Hellman, is a method by which two nodes can agree on a key they can use to encrypt a conversation. The protocol for key exchange Diffie-Hellman, based on a function of the form $K = W^x \mod P$ with the first P and W<P.
This function is very easy to calculate, knowledge of K does not allow to deduce X easily.
This function is public, and the values of W and P.
The two VANET network users Mohammed and Ayoub each choose a secret number used as the exponent and proceeds as follows:
1. Ayoub chooses a number that will remain his secret, say A.
2. Mohammed chooses a number that will remain his secret, say B.
3. Ayoub and Mohammed want to exchange the secret key, which is $S = W^{B.A} \mod P$, but they do not yet know, since everyone knows that A or B, but not both.
4. Ayoub applies to A the one-way function, α is the result: $\alpha = W^A \mod P$
5. Mohammed applies to B the one-way function, β is the result: $\beta = W^B \mod P$
6. Ayoub sends α to Mohammed, and Mohammed sends β, as shown by, they may be known to the whole world without the secret of Ayoub and Mohammed is disclosed.

7. Ayoub received β and calculates $\beta^A \bmod P$ (that is to say in passing by $(W^B)^A \bmod P$ but he does not know B): $S = \beta^A \bmod P$

8. Mohammed received α and computes $\alpha^B \bmod P$ (that is to say in passing by $(W^A)^B \bmod P$, but he does not know A): $S = \alpha^B \bmod P$

Mohammed and Ayoub get to the end of their respective calculations the same number that has never been exposed to the sight of prying: the S key.

Figure 1 presents the processes key exchange Diffie Hellman.

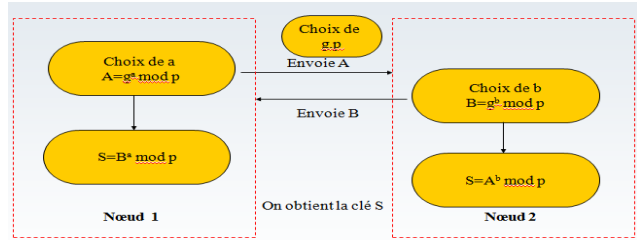

Figure 1: Diffie Hellman Key Exchange

- Relationship Type Master-Slave (Duckling Security Policy Model):

The work of Ross Anderson and Frank Stajano part of the concept of ubiquitous computing, ubiquitous computing. In this model, most objects around us are destined to receive a processor to enable them to make records of measurement or to render services to other objects such as the master PDA (Personal Digital Assistant) a doctor. Communication between objects is established via the radio channel. The authentication model developed by Ross Anderson and Frank Stajano [6], The Duckling Security Policy Model, is based on a relationship of master-slave type. At its first use, an object must be marked, imprinting, by the owner. During this operation a secret key is exchanged between the two entities via a secure channel.

The third research orientation for authentication in wireless ad hoc, based on public key cryptography and seeks to overcome the need for a central entity certification:

-The public key infrastructure self-organized: Model proposed by Hubaux et al. [7].

These solutions and cryptographic hardware and software implementations to implement in VANET networks must meet the requirements and security challenges, such as confidentiality, source authentication or mutual authentication and message integrity.

## 3. CRYPTOGRAPHIC SYSTEM AND EXPERIMENTATION

### 3.1. Advanced Encryption Standard (AES)

It comes from an international call for applications launched in January 1997 and has received 15 proposals. Of these 15 algorithms, five were selected for further evaluation in April 1999: MARS, RC6, Rijndael, Serpent, and Twofish. After this assessment, it was finally the candidate Rijndael, named after its two designers Joan Daemen and Vincent Rijmen (both Belgian nationality) who has been chosen [9,10]. These two experts in cryptography were already authors of another algorithm: Square. AES is a subset of Rijndael: it only works with blocks of 128 bits, whereas Rijndael offers block sizes and keys that are multiples of 32 (between 128 and 256 bits).

In so doing the AES replaces the DES (chosen as standard in the 1970s) which today became obsolete, because it used only 56-bit keys. The AES has been adopted by NIST (National Institute of Standards and Technology) in 2001 [9, 10]. Moreover, its use is very convenient because it uses little memory and is not based on a Feistel scheme, its complexity is lower and it is easier to implement.

International Journal of Distributed and Parallel Systems (IJDPS) Vol.2, No.5, September 2011

The algorithm takes as input a block of 128 bits (16 bytes), the key is 128, 192 or 256 bits. The 16 input bytes are swapped according to a predefined table. These bytes are then placed in a 4x4 matrix components and lines are rotated to the right. The increment for the rotation varies with the number line. A linear transformation is then applied to the matrix, it consists of a binary multiplication of each element of the matrix with polynomials from an auxiliary matrix, this increase is subject to special rules as GF (28) (Galois group finite) [9,10]. The linear transformation ensures a better distribution (propagation of bits in the structure) on several laps. Finally, an XOR between the matrix and another matrix provides an intermediate matrix. These different operations are repeated several times and set a "tower". For a key of 128, 192 or 256, AES requires respectively 10, 12 or 14 towers.

### 3.2. Message Digest 5

The hash is used to produce a message digest is a unique reduced representation of the complete message. The hash algorithms are one-way encryption algorithms, so it is impossible to recover the original message from the digest. MD5 is a hash function block [11]. That is to say that cutting the chopping block messages of fixed size and is working on a block at a time. If the message size chopping is not a multiple of block size, it will be completed (this is the padding operation) until a complete block. We chose MD5 because it works on blocks of 512bit (64 bytes) and produces a digest of 128 bits.

### 3.3. Comparison of execution time

Our goal is to measure and compare the execution time of the AES and Blowfish to prove that the AES and faster and more responsive to our needs for security in the context of vehicular ad hoc networks.

We will use to measure the execution time of AES and Blowfish encryption algorithms that we implemented in Java System.nanoTime function () which returns the current value of the system clock as accurate as available, in nanoseconds.

/ / Example of the function System.nanoTime ()

Long t1, t2;

System.nanoTime t1 = ();

/ / code that wants to calculate the execution time

System.nanoTime t2 = ();

    System.out.println ("time decryption:" + (t2-t1));

The comparison is to the lengths of 128.256 and message of 512 bytes.

Table 1 shows the execution time depending on the size of the message.

International Journal of Distributed and Parallel Systems (IJDPS) Vol.2, No.5, September 2011

| Taille du message en octet / AES | 128 | 256 | 512 |
|---|---|---|---|
| Temps chiffrement en nanosecondes | 6383481 | 7166724 | 8684963 |
| Temps Déchiffrement en nanosecondes | 2032531 | 2975913 | 5500665 |

| Taille du message en octet / Blowfish | 128 | 256 | 512 |
|---|---|---|---|
| Temps chiffrement en nanosecondes | 16396774 | 18365661 | 15570931 |
| Temps Déchiffrement en nanosecondes | 954160 | 1023451 | 1124051 |

Table 1: Execution Time of AES and Blowfish

### 3.4 Form of GPSR protocol packets

In this section we present our GPSR protocol specification by introducing some changes to the format of the packet GPSR [4].
GPSR protocol specification:

| Version | Command | Reserved | Packet length |
|---|---|---|---|
| Position of the source ||||
| Position Perimeter (optional) ||||
| Position of the destination ||||
| Edges ||||
| Port || Reserved ||
| Data ||||
| Authentification ||||

Fields:
 - Version: protocol version 4 bits
 - Command: message type 4 bits
 1: request beacon: 0001
 2: response beacon: 0010
 3: Data forwarding mode"Greedy":0011
 4: Data Forwarding mode "Perimeter ":0100
 - Reserved: 8 bits
 - Length of the package: 16 bits
 - Position of the source: 64 bits
 - Position perimeter: 64 bits
 - Position of the destination: 64bit
 - Edge position between P1 and P2:128bits
 - Port: 16 bits

100

- Data (max 250 byte)
- Authentication

### 3.5 The message signature GPSR

The signature is calculated on the message body, and is distributed in the form of a message field, called authentication. This solution allows controlling the integrity of one hop, but an authentication message can easily determine the origin of false information. We propose to use an authentication between each two neighbour by agreeing on a secret key: A method of Diffie-Hellman, AES then be used to sign the hash MD5 of the package, and then put it in the field authentication. The recipient verifies the message by decrypting the field and comparing the hash of the received message with that of the original message.

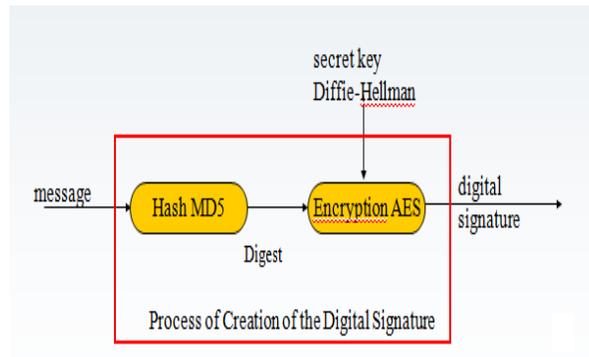

Figure 2: Process of Creating the Digital Signature

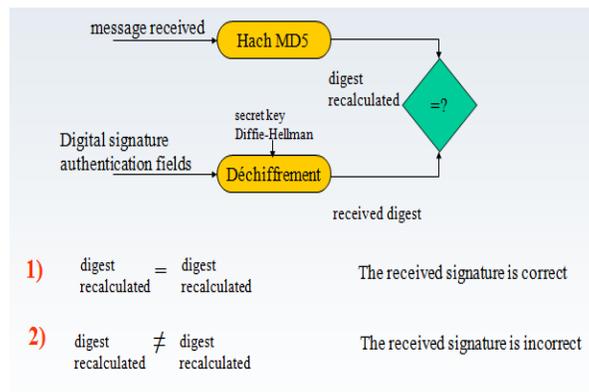

Figure 3: Process for checking the digital signature.

### 3.6 Secured GPSR:

To ensure the security of messages during the transfer GPSR have the verification between each two nodes.
Principle:
- Generate a shared secret key between two neighbor when the neighbor discovered using Diffie-Hellman (n (n-1) / 2 keys), this key will be used as AES encryption key.
- Application of a hash function on the GPSR packet header. (Integrity Management)
- Implementation of the AES algorithm to generate the signature to be added to the package GPSR.

101

## 4. CONCLUSIONS

Vehicular Ad-Hoc Network (VANET) present some particularities such as the mobility , the independence to any infrastructure and it has an uncommon routing layer ,thus new routing protocols should be designed .however , difficult challenge exist in the design of VANET secure routing protocols, since they should not only be robust against various attacks but also efficient in terms of performances .

In this work we have chosen to study the routing protocol GPSR and we are interested in the concept of digital signature as a security solution, this concept is particularly suited to the mobile nature of vehicular ad hoc networks.

We have at first given an overview on possible attacks against the ad hoc routing protocol that we will take into account in the design of our secure routing mechanisms, secondly we have presented a state of art of security mechanism used for parked the authenticity of routing elements .

At last, we gave our contribution by suggesting a solution to secure GPSR. The solution is to add a digital signature symmetric on the routing packet exchanged on the network, which is protection against intrusions into the routing protocol GPSR. This signature is short and quick and a lower computational complexity makes the proposed security architecture best suited to the reality of a vehicular ad hoc protocol.

We hope for further develop this work along two axes: the first axe is the finalization of the software platform of our protocol, the second axe is the implementation of an intrusion detection system (IDS) to detect any violation the security policy on vehicular ad hoc network.

The first axis is a natural extension of this work. We want to put very quickly a fully functional version of our protocol SGPSR available to users. We want also adapted our protocol to other constraints environment such as sensor networks. The second axis will be an experimental study of Bayesian networks to detect intrusions on Vehicular ad hoc networks.

## REFERENCES

[1] Sofiane Khalfallah, Moez Jerbi, Mohamed Oussama Cherif , Sidi-Mohammed Senouci , Bertrand Ducourthial, Expérimentations des communications inter-véhicules, Colloque Francophone sur l'Ingénierie des Protocoles (CFIP), Les Arcs : France (2008)

[2] ADJIDO Idjiwa, BENAMARA Radhouane, BENZIMRA Rebecca, GIRAUD Laurent, Protocole de routage ad hoc sécurisé dans une architecture clusterisée.

[3] Valerie Gayraud, Lout_ Nuaymi, Francis Dupont, Sylvain Gombault, Bruno Tharon, La Sécurité dans les Réseaux Sans Fil Ad Hoc, Actes du symposium SSTIC03

[4] B. Karp and H. T. Kung, « Gpsr: Greedy perimeter stateless routing for wireless networks », in Proceedings of ACM/IEEE MOBICOM'00, Boston, USA, Aout 2000.

[5] Juan-Carlos Ruiz, Jesús Friginal, David de-Andrés, Pedro Gil , ''Black Hole Attack Injection in Ad hoc Networks'',Fault Tolerance Systems Group (GSTF), Instituto de las TIC Avanzadas (ITACA) Universidad Politécnica de Valencia, Campus de Vera s/n, E-46022, Valencia, Spain.

[6] Frank Stajano and Ross Anderson, The Resurrecting Duckling : Security Issues for Ad-hoc Wireless Networks, 7th International Workshop on Security Protocols, 1999.

[7] Hubaux, J. P., Buttyan, L. and Capkun, S., The Quest for Security in Mobile Ad Hoc Networks, Proceedings of ACM Symposium on Mobile Ad Hoc Networking and Computing (MobiHOC), Long Beach, CA, 2001.

[8] ALIOUANE Lynda , BADACHE Nadjib,L'Authentification dans les Réseaux Ad Hoc.

[9] National institute of standards and technology (NIST),Advanced Encryption standard (AES) Conference, (Rome, Italy), March 1999.

International Journal of Distributed and Parallel Systems (IJDPS) Vol.2, No.5, September 2011

[10]     National institute of standards and technology (NIST),Advanced Encryption standard (AES), Federal Information Processing Standards (FIPS) publication197,2001.

[11]     Simon Marechal , Etat de l'art sur le cassage de mots de passe, Actes du symposium SSTIC07 ,2007.

[12]      Hassiba-asmaa Adnane , la confiance dans le protocole de routage ad hoc :étude du protocole OLSR,2008 .

[13]     Manel Guerrero Zapata, Internet draft, "Secure Ad hoc On-Demand Distance Vector (SAODV) Routing", 15 septembre 2005

[14]     Yih-Chun Hu, Adrian Perrig, David B. Johnson, "Packet Leashes: A Defense against Wormhole Attacks in Wireless Networks" ,infocom 2003.

[15]     M. Bechler, H.-J. Hof, D. Kraft, F. Pählke, L. Wolf,  "A Cluster-Based Security Architecture for Ad Hoc Networks" ,infocom 2004.

[16]     Yih-Chun Hu, Adrian Perrig, David B. Johnson, "Rushing Attacks and  Defense in Wireless Ad Hoc Network Routing Protocols" ,2003.

[17]     Panagiotis Papadimitratos, Zygmunt J. Haas, Prince Samar, " The Secure Routing Protocol (SRP) for Ad Hoc Networks" , December 2002

[18]     Yih-Chun Hu, Adrian Perrig, David B. Johnson, "Ariadne: A Secure On Demand Routing Protocol for Ad Hoc Networks"

**Authors**

Bouabid  El ouahidi
The Head of computer sciences department and the Network and Data Mining laboratory of the faculty of sciences, Mohamed V University. Obtained a PhD in Computer Sciences from the University of Caen at France in 1992 and a PhD in Computer Sciences from the Mohamed V University at Morocco in 2002. His current interests include developing specification and design techniques for use within Intelligent Network.

Oussama Mohamed Reda
Member of Network and Data Mining laboratory and professor in the faculty of sciences of Mohamed V University in Rabat where he has obtained his PhD in open distributed systems in 2009. His current interests include Mobile computing and networking, intelligent transportation and business intelligence systems.
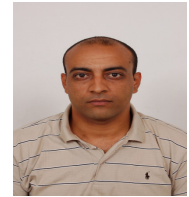

Mohammed ERRITALI
PhD student and member of the Network and Data Mining laboratory of the faculty of sciences, Mohamed V University, Rabat. Obtained a master's degree in business intelligence from the faculty of science and technology, Beni Mellal at Morocco in 2010.
His current interests include developing specification and design techniques for use within Intelligent Network, data mining and cryptography.
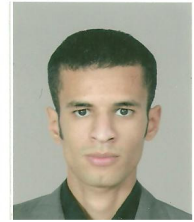